\documentclass{PoS}

\title{Jet substructure in high-energy hadron collisions}

\ShortTitle{Jet substructure in high-energy hadron collisions}

\author{\speaker{Felix Ringer}\\ 
        Nuclear Science Division, Lawrence Berkeley National Laboratory, Berkeley, CA 94720, USA \\
        E-mail: \email{fmringer@lbl.gov}}

\abstract{In the past years significant progress has been made toward achieving a quantitative understanding of jets and their substructure in high-energy proton-proton collisions from first principles in QCD. Precise measurements have become available from the experimental collaborations at the LHC and RHIC allowing a direct comparison of theoretical calculations and data. These developments make it possible to use jet substructure observables as precision probes in heavy-ion collisions. The radiation pattern inside jets contains valuable information about the hot and dense QCD medium which can be investigated using jet substructure techniques. By studying the soft sensitivity of the different observables it is possible to obtain important insights into the interaction of hard probes with the quark gluon plasma.}

\FullConference{International Conference on Hard and Electromagnetic Probes of High-Energy Nuclear Collisions\\
		30 September - 5 October 2018\\
		Aix-Les-Bains, Savoie, France}

\begin{document}

\section{Introduction}

In the past decade jet substructure observables have become an important tool at the LHC and RHIC. Applications of jet substructure techniques include quark/gluon discrimination, searches of boosted objects such as $W$/$Z$/top quarks and physics beyond the standard model~\cite{Butterworth:2008iy}. In recent years significant progress has been made toward calculating jet substructure observables in proton-proton collisions from first principles in QCD making the precise measurement of cross section of great interest. Advances both from the theoretical and experimental side allow for a quantitative understanding of the radiation pattern observed inside jets and utilize jet substructure as a tool to probe various interesting physics aspects. In particular, jet substructure observables are now used to achieve a better understanding of jets that probe the hot and dense quark gluon plasma (QGP) created in heavy-ion collisions. The modified substructure of jets in heavy-ion collisions relative to the proton-proton baseline contains information about the nature of the created plasma. Jet substructure observables currently provide new challenges and opportunities for model calculations. Being able to control the sensitivity of jet substructure observables to soft radiation it is possible to learn about different aspects of how hard probes interact with the medium which provides stringent constraints for models of the QGP. In order to allow for a one-to-one comparison between theory and experiment it is crucial to understand jet substructure observables at the level of factorization theorems within perturbative QCD which we review in this work. \\

In section~\ref{sec:pp}, we will first review the status and recent progress that has been made in jet substructure calculations in proton-proton collisions. In section~\ref{sec:hi} we will focus on applications of jet substructure techniques to the more complicated environment of heavy-ion collisions and we draw conclusions and present an outlook in section~\ref{sec:con}.

\section{Proton-proton collisions \label{sec:pp}}

There is a large range of jet substructure observables which have been proposed and measured in proton-proton and heavy-ion collisions. For heavy-ion collisions as discussed in section~\ref{sec:hi}, a useful criterion to categorize the different observables is their sensitivity to soft physics. We consider inclusive jet production $pp\to{\rm jet}+X$. The first category of jet substructure observables is only sensitive to final state collinear radiation within the phenomenologically relevant kinematical range. In particular, the required factorization theorem does not involve a soft function. Examples of such observables are the longitudinal momentum distribution of hadrons~\cite{Procura:2009vm} or inclusive subjets~\cite{Kang:2017mda}. These observables map out the longitudinal momentum structure of jets where the momentum of each hadron or subjet is measured and and projected onto the jet axis. The momentum fraction $z_h=p_T^h/p_T$ measures how much of the total jet $p_T$ is carried by the individual hadrons $p_T^h$. The factorization theorem for such observables only depends on parton distribution functions $f_{a,b}$ (PDFs), hard-scattering functions $H_{ab}^c$ and collinear jet functions ${\cal G}_c$. Schematically, we have the structure~\cite{Kaufmann:2015hma,Kang:2016ehg}
\begin{equation}\label{eq:fact}
\frac{d\sigma^{pp\to{\rm jet}+X} }{dp_T d\eta dz_h} =\sum_{abc}f_a\otimes f_b\otimes H_{ab}^c \otimes {\cal G}_c(z_h)\,,
\end{equation}
where $p_T$ and $\eta$ are the jet's transverse momentum and rapidity and the $\otimes$ denote appropriate integrals over the involved partonic momentum fractions. Note that the dependence on the jet substructure variable $z_h$ only appears in the jet function ${\cal G}_c$ which can be calculated at fixed order in perturbation theory. The simplicity of the factorization theorem in Eq.~(\ref{eq:fact}) allows for a direct comparison of data and calculations from first principles in QCD. For example, it allows for a direct extraction of the non-perturbative fragmentation functions within a global analysis~\cite{Anderle:2017cgl} which gives important insights into the QCD hadronization mechanism. Other jet substructure observables can be calculated analogously using the factorization in Eq.~(\ref{eq:fact}) except that the appropriate jet functions ${\cal G}_c$ need to be calculated.

\begin{figure}[t!]
\begin{center}
\includegraphics[width=0.975\textwidth]{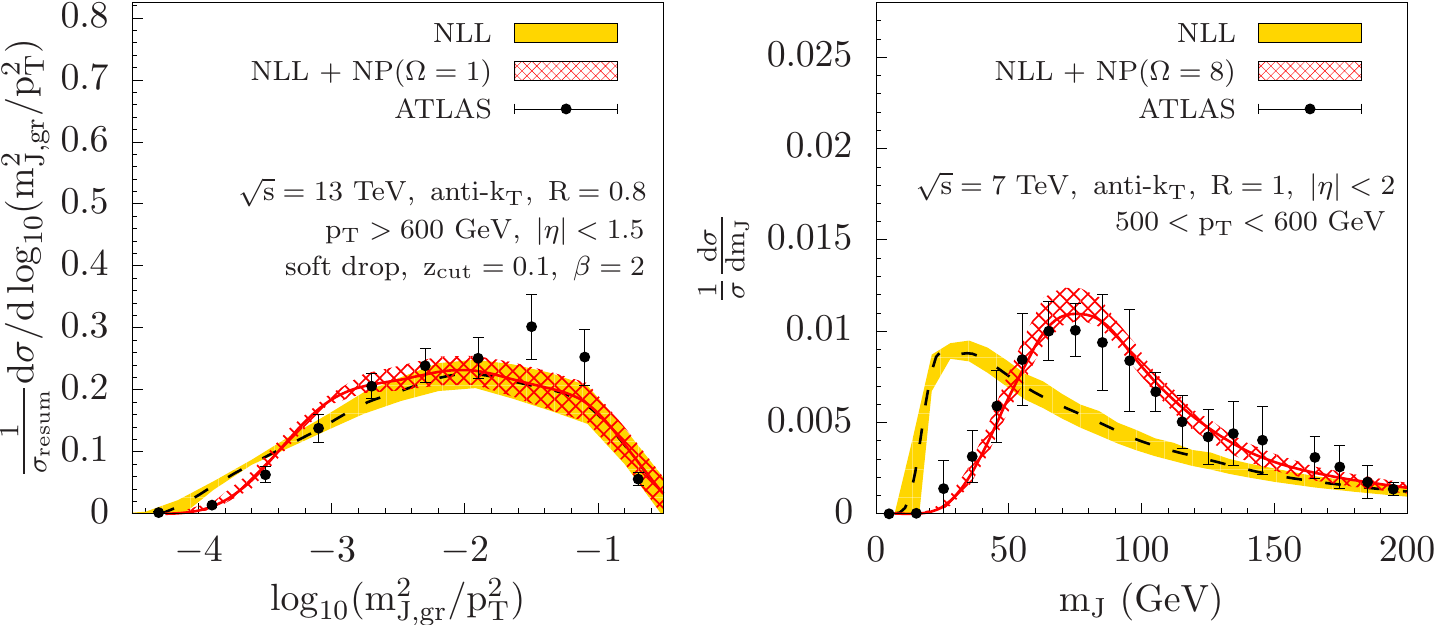}
\caption{ 
The soft drop groomed (left) and ungroomed (right) jet mass distribution~\cite{Kang:2018jwa} compared to the ATLAS data of~\cite{ATLAS:2012am,Aaboud:2017qwh}.
\label{fig:1}}   
\end{center}
\end{figure}

In order to understand also the transverse momentum structure of jets or the shape of jets, fixed order calculations of the relevant jet functions ${\cal G}_c$ are typically not sufficient. The presence of large logarithms may spoil the convergence of the perturbative expansion in $\alpha_s$. For example, in order to describe the jet mass distribution $m_J$ in the phenomenologically relevant kinematic region, large logarithms of the form $\alpha_s^n\ln^{2n}(m_J/p_T)$ need to be taken into account to all orders in the strong coupling constant. This can be achieved by refactorizing the jet function in terms of hard, collinear and soft functions ${\cal G}_c\sim {\cal H}_{cd}\, C_d \otimes S_d$, where the dependence on $m_J$ appears both in the collinear $C_d$ and the soft function $S_d$. The different functions all satisfy renormalization group (RG) equations which can be used to resum the corresponding large logarithms. Observables which require such a factorization are the second category of observables discussed here. Besides the jet mass they include for example the transverse momentum structure of jets, jet angularities, the jet energy profile as well as observables that probe the multi-prong structure of jets such as $N$-subjettiness~\cite{Thaler:2010tr} or $D_2$~\cite{Larkoski:2017cqq}. These observables are generally sensitive to very soft scales and thus introduce a dependence on initial state radiation (ISR) and non-global structures and they require a model of the underlying event (UE)/multi parton interactions (MPI) as well as hadronization. While the increased soft sensitivity is interesting in its own right, it makes a direct comparison of data with first principles calculations in QCD more difficult. As an example, we show a comparison of theoretical results of the jet mass distribution~\cite{Kang:2018jwa} (dashed black, yellow band) and ATLAS data~\cite{ATLAS:2012am} on the right side of Fig.~\ref{fig:1}. The resummation of large logarithms here is performed at next-to-leading logarithmic (NLL) accuracy. The data was taken at a center of mass (CM) energy of $\sqrt{s}=7$~TeV where jets are reconstructed with $500<p_T<600$~GeV and $|\eta|<2$ using the anti-k$_T$ algorithm with $R=1$. The peak of the purely perturbative calculation is around $m_J\approx 30$~GeV whereas the peak of the data is approximately at a jet mass value of $m_J\approx 70$~GeV. This large discrepancy is due to the corrections listed above, in particular due to hadronization and the underlying event. By using a non-perturbative shape function model which is to be convolved with the perturbative result, a good description of the ATLAS data can nevertheless be achieved as shown in red (hatched band) in Fig.~\ref{fig:1}. The shape function depends on a single parameter $\Omega$ as indicated in the figure and it is the same for the different jet $p_T$ bins. When these type of jet substructure observables are measured in the more complicated heavy-ion environment it is thus difficult to disentangle which part of the cross section is modified. In addition, in heavy-ion collisions it is important that the measured jet substructure observable is directly related to the fragmenting parton that produces the jet and that contains the information about the QGP. However, observables that are very soft sensitive also contain information about the soft radiation in the entire event which makes it challenging to pin down where the modification of jet substructure cross sections in heavy-ion collisions is coming from.

The third category discussed here are jet substructure observables that involve a jet grooming algorithm. Various grooming techniques have been developed in the literature which are designed to systematically remove soft radiation such that the observed jet is reduced to its hard, collinear core. The most frequently used procedure now is soft drop declustering~\cite{Larkoski:2014wba}. The first step is to recluster a given jet with the Cambridge/Aachen algorithm which yields an angular ordered clustering tree. Afterwards soft branches are removed from the jet by iterative declustering until the criterion $z>z_{\rm cut}(\Delta R_{12}/R)^\beta$ is satisfied. Here $z$ is the energy fraction of the softer branch and $\Delta R_{12}$ is the geometric distance of the two branches in the $\eta-\phi$ plane for a given declustering step. The parameters $z_{\rm cut}$ and $\beta$ can be adjusted to control the sensitivity to wide-angle soft radiation. The jet substructure measurement is then performed on all the remaining particles in the groomed jet which may include all of the observables discussed above. Note that the ungroomed case is obtained in the limit $\beta\to\infty$. In addition, there are observables such as the groomed radius $R_g$ or the momentum sharing fraction $z_g$~\cite{Larkoski:2015lea} which are only defined for groomed jets in the sense that they do not have an analogue in the ungroomed limit. In order to illustrate the impact of the soft drop grooming algorithm at the level of the factorization theorem and to study the corresponding soft sensitivity, we consider again the jet mass distribution. Theoretical calculations of the soft drop groomed jet mass spectrum have been performed in~\cite{Frye:2016aiz,Marzani:2017mva,Kang:2018jwa}. The hard and collinear functions remain the same as for the ungroomed case. However, the soft function $S^{\rm gr}_d$ that contains the dependence on the jet mass now also depends on the grooming parameters. Moreover, there is an additional soft function $S^{\not\in{\rm gr}}_d$ that takes into account soft radiation which fails the soft drop criterion. Thus, the relevant refactorized jet function is given by ${\cal G}_c\sim {\cal H}_{cd}\,S_d^{\not\in {\rm gr}}\, C_d\otimes S^{\rm gr}_d$. The associated RG equations can now be used to also resum logarithms in the grooming parameter $z_{\rm cut}$. On the left hand side of Fig.~\ref{fig:1}, we show the comparison of the theoretical results of~\cite{Kang:2018jwa} and the ATLAS data of~\cite{Aaboud:2017qwh} for the soft drop groomed jet mass distribution $m_{\rm J,gr}$ with $z_{\rm cut}=0.1$ and $\beta=2$. The data corresponds to jets with $p_T>600$~GeV and $|\eta|<1.5$ at a CM energy of $\sqrt{s}=13$~TeV. Different than the ungroomed case, we already find a relatively good agreement of the data and the purely perturbative result and only a small hadronization correction is needed. Note that the slight disagreement at large values of the groomed jet mass is due to an additional cut $p_{T1}/p_{T2}<1.5$ applied to the ATLAS data which is not taken into account in the theoretical calculation. At sufficiently high jet mass values the radiation is sufficiently energetic such that the grooming procedure drops out. This transition occurs at $m_J^2=z_{\rm cut} R^2 p_T^2$. The good agreement between theory and data confirms that the grooming procedure indeed reliably removes soft radiation from the jet allowing a direct one-to-one comparison of data and first principles calculations in QCD. In addition, these findings imply that we may directly associate the measured mass spectrum with the parton initiating the jet which contains the desired information about the QGP in heavy-ion collisions after traversing the hot and dense QCD medium.

Besides the fact that groomed jet substructure observables constitute systematic and controllable probes of the QGP in heavy-ion collisions, they allow for a range of precision QCD studies in proton-proton collisions. For example, it has recently been proposed that it will be possible to extract the QCD strong coupling constant $\alpha_s$ from groomed jet substructure observables such as the jet mass~\cite{Bendavid:2018nar}.

\section{Heavy-ion collisions \label{sec:hi}}

In heavy-ion collisions at the LHC and RHIC, the QGP can be produced which is predicted to have existed during the early stages of our universe. Highly energetic particles and jets traverse the hot and dense QCD medium and, hence, provide an excellent tool to probe this new state of matter. In the past years, the proliferation of jet substructure observables has provided new differential probes of the QGP providing novel constraints for model calculation of the QCD medium. Jet cross sections in heavy-ion collisions are conveniently studied by comparing the cross section to the corresponding result in proton-proton collisions. The nuclear modification factor $R_{AA}$ is defined as
\begin{equation}
R_{AA}=\frac{d\sigma^{AA\to{\rm jet}+X}}{\langle N_{\rm coll}\rangle d\sigma^{pp\to{\rm jet}+X}}\,,
\end{equation}
where $\langle N_{\rm coll}\rangle$ is the average number of binary nucleon-nucleon collisions. The nuclear modification factor and its centrality dependence have been studied for a range of observables including the distribution of hadrons inside jets, the jet energy profile, the (groomed) jet mass distribution and the momentum sharing fraction $z_g$. In this work, we will review different approaches that can be used in order to extract information about the QGP and in particular to better understand its interaction with hard probes using jet substructure techniques. First, it is possible to design suitable jet substructure observables that directly test interesting aspects of the in-medium interactions with hard probes. For example, the QCD splitting functions can be probed using the momentum sharing variable $z_g$ that can be calculated as a Sudakov safe observable within the soft drop grooming procedure~\cite{Larkoski:2015lea}. Alternatively, inclusive subjets~\cite{Kang:2017mda} provide a similar handle on the splitting function and jet functions in the medium. An interesting approach to probe the dynamics of how jets get modified in heavy-ion collisions is to use multi-prong jet substructure observables such as $N$-subjettiness or $D_2$ including soft drop grooming. Multi-prong jet substructure observables were initially designed to discriminate between one-pronged QCD jets and two- or three-pronged jets that originate from boosted heavy objects such as $W$, $Z$ or top quarks. In the context of heavy-ion collisions these observables can be used to separate the measured jet sample into jets that have one or several hard cores. If the medium can resolve both hard cores independently, multi-pronged jets are expected to lose more energy. This effect is known as color decoherence which can be studied quantitatively using jet substructure observables. In~\cite{Zardoshti:2017yiy}, the $N$-subjettiness observable was used to investigate this effect. In the future it will be interesting to study these observables in more detail including grooming techniques. For example, in~\cite{Larkoski:2017cqq} the $D_2$ observable was calculated analytically in proton-proton collisions using the soft drop grooming algorithm. Recently, it was also proposed to measure heavy quark properties using interative jet declustering~\cite{Cunqueiro:2018jbh}.

Second, starting from QCD factorization in proton-proton collisions it is possible to learn important lessons about how the QGP couples to hard probes. Interestingly, it was found that the jet mass spectrum is largely unmodified within the experimental uncertainties~\cite{Acharya:2017goa,Sirunyan:2018gct,ATLAS:2018jsv}. However, other observables such as the closely related jet broadening/girth show a large and non-trivial modification pattern~\cite{Acharya:2018uvf}. Given their close relation it will be interesting to measure (groomed) jet angularities~\cite{Kang:2018vgn} which allow a smooth interpolation between different jet substructure observables and thus achieve a better understanding of which observables are modified and why. A possibility to control the soft sensitivity of jet substructure observables is to use different choices of the jet axis. The standard jet axis is obtained by adding the four-momenta of the particles that are clustered together. Instead, the winner-take-all axis is aligned with the more energetic particle at each clustering step. Such a recoil free axis affects the structure of QCD factorization theorems. Recently phenomenological studies using the winner-take-all jet axis were performed in~\cite{Neill:2018wtk}. Measuring the same observables but with different jet axis definitions can be used as a direct probe of the soft sensitivity of in-medium interactions.

The third approach is to directly construct microscopic models of the QGP and its interaction with highly energetic jets. These approaches can benefit from the other two discussed above in order to reduce their model dependence. Various approaches have been considered in the literature. In~\cite{Chien:2016led,Li:2017wwc} an effective field theory approach was used that incorporates the interaction of highly energetic partons with the medium at the level of the QCD Lagrangian. From the obtained in-medium splitting functions, the modification of the $z_g$ spectrum was calculated. A different approach was taken in~\cite{Mehtar-Tani:2016aco}, where coherent and incoherent energy loss was addressed. Both of these approaches rely largely on analytical methods in order to calculate the modified jet spectra. Several other approaches rely on parton shower event generators that model a fully exclusive final state. In order to incorporate in-medium effects various models have been studied. For example, in~\cite{Luo:2018pto}, a linearized Boltzmann transport code was included in order to calculate the jet energy profile for photon tagged jets.  In~\cite{Tachibana:2017syd}, the interactions of the parton shower with the medium were modeled including collisional and radiative processes and the medium is modeled using a hydrodynamical simulation. The jet mass and jet energy profile were calculated in~\cite{Park:2018acg} using medium induced energy loss and hydrodynamical simulations. See also~\cite{KunnawalkamElayavalli:2017hxo} where a range of jet substructure observables in heavy-ion collisions were calculated. Several of these studies discuss when a particle should to be considered as part of the medium or the jet and they emphasize the relevance of medium response in order to describe jet substructure observables. Finally, in~\cite{Brewer:2017fqy} a holographic approach was developed in order to study the jet energy profile in heavy-ion collisions.

\section{Conclusions\label{sec:con}}

Our understanding of jet substructure observables has increased rapidly in the past years. High precision experimental results have become available from the LHC and RHIC. From the theory side an understanding of of jet substructure observables has been achieved from first principles in QCD allowing a meaningful one-to-one comparison between theory and data. While some observables still require a systematic extension beyond leading oder/leading-logarithmic accuracy, other observables like the jet mass distribution have reached an unprecedented level of precision including the resummation of logarithms at next-to-next-to leading logarithmic order. First extractions of non-perturbative quantities from jet substructure data have been performed such as the extraction of fragmentation functions. A future goal will be to achieve the precision sufficient for a competitive extraction of the QCD strong coupling constant from jet substructure data. Given the fact that jet substructure observables are well calibrated tools in proton-proton collisions, they can be used to unravel new aspects of the QGP created in heavy-ion collisions. In the future it will be interesting to measure observables such as (groomed) jet angularities over a wide kinematic range in order to systematically increase our understanding of how jets get modified in the QCD medium. A possibility to control the soft sensitivity of jet substructure observables is to compare the results using different jet axes. In addition, the detailed measurement of observables sensitive to the multi-prong nature of jets such as $N$-subjettiness or $D_2$ will be very valuable. At the same time it will be crucial to keep a close connection to first principles calculations performed in proton-proton collisions using QCD factorization. This way it will be possible to achieve a better understanding of jets in heavy-ion collisions and suitable jet substructure measurements can provide guidance for constructing microscopic models of the QGP and its interaction with hard probes.

\section*{Acknowledgements}

I would like to thank Zhong-Bo Kang, Kyle Lee, Xin-Nian Wang and Feng Yuan for helpful discussions regarding this writeup. This work is supported by the Department of Energy under Contract No. DE-AC0205CH11231, and the LDRD Program of Lawrence Berkeley National Laboratory.

\end{document}